\documentstyle[preprint,revtex]{aps}
\tightenlines
\begin{document}

\begin{title}
Flux flow dissipation in superconductors with short
coherence length \\
\end{title}

\author{F. Guinea \\}
\begin{instit}
	Instituto de Ciencia de Materiales (CSIC),
	Universidad Aut\'onoma de Madrid,\\
	E-28049 Madrid, Spain
\end{instit}

\author{Yu. Pogorelov}
\begin{instit}
	Departamento de Fisica de la Materia Condensada,
	Universidad Aut\'onoma de Madrid, \\
	E-28049 Madrid, Spain
\end{instit}

\receipt{}

\begin{abstract}

In superconductors where the coherence length is comparable to
the Fermi wavelength, the electronic levels within
a vortex core are quantized, and separated by energies
of the order of the superconducting gap.
The absence of a continuum modifies significantly the energy
dissipation due to the motion of vortices. At zero temperature,
dissipation is suppressed, and the system shows non ohmic behavior.
The I-V characteristics show two distinct regimes. Below
a given electric field,  I is strongly dependent on V.
Flux flow resistivity also shows a marked dependence on
temperature, and reaches
conventional values near $T_c$.

\end{abstract}


\narrowtext

Flux flow dissipation is ubiquitous in type-II superconductors.
The standard explanation of this effect is that, when
vortices move, part of the electrical current flows
through the core\cite{BS,NV} (see also\cite{deGennes}).
The core region is assumed to be similar to a normal metal.
Dissipation in the core is described by means of processes
indistinguishable from those which take place within a normal
metal
(note also that at fields close to $H_{c_2}$, or temperatures close to $T_c$,
additional dissipation arises from quasiparticles everywhere
in the system\cite{Schmid,Maki}).

The basic premise of this argument is that, inside the core, a
quasicontinuum of bound states exists.
The motion of the vortices with respect
to the impurities in the lattice gives rise to transitions
for any velocity or energy. This picture is supported
by detailed calculations for the low lying states
within a vortex core\cite{deGennes2,Schluter}.

Most of the Cu based high-$T_c$ superconductors have
coherence lengths at zero temperature very short,
comparable to the lattice constant and to
the Fermi wavelength. In this regime, usual calculations
for the states at the vertex core are not applicable.
It has been argued that the states within the core should be
quantized, and separated by energies comparable to
the superconducting gap. Recent observations
confirm this picture\cite{states}.

The standard models of flux flow resistivity do not
describe microscopically the processes which give
rise to dissipation, which are taken to be the same as
in the normal state.
In normal metals, there is a continuum of
electron-hole excitations. The density of states of
these modes is proportional to its energy. This assumption
suffices to obtain ohmic dissipation due to impurity
scattering\cite{Ashcroft,Guinea}.

In the absence of a continuum, this picture needs
to be reformulated. We will assume, following the
standard work on flux flow resistivity that,
when vortices move, some current flows through their
core. As in conventional superconductors,
dissipation takes place through the excitation
of transitions between states within the core.
We study these processes
in a frame of reference where the
vortices are at rest, and lattice impurities are
displaced towards them. We will consider impurity scattering
only, although the scheme can easily be generalized
to other kinds of disturbances. We assume that
impurities are the most efficient mechanism at
low temperatures. Initially, each impurity is described
by the Fourier transform of the potential,
$\sum_q V_q e^{i \vec{q} \vec{r}}$. In the moving frame
of reference, the position of a given impurity
with respect to a static vortex located at the origin is
$\vec{r} ( t ) = \vec{r_0} + \vec{v} t$, where $\vec{v}$
is the relative velocity of the impurities.
Hence, the impurities give rise
to a time dependent potential, $\sum_q V_q e^{i \vec{q}
\vec{r_0}} e^{i \vec{v} \vec{q} t}$. Dissipation is due to transitions
induced by this potential.

This picture is standard in the study of atom-solid collisions
\cite{Echenique}. Commonly, the excitation spectrum of the material,
where energy is dissipated,
is described in terms of response functions. Using this picture
for an homogeneous superconductor, we obtain that dissipation sets
in when $\hbar q_{max} v_s = \Delta$, where $\Delta$
is the gap of the superconductor, $q_{max}$ is the
maximum wavevector for an excitation at low energies
and $v_s$ is the velocity of the superfluid.
Setting the free electron value $q_{max} = 2 k_F$,
we recover the known result for the critical
current in a superconductor, in the absence of a
magnetic field\cite{deGennes}.

A knowledge of the full response function of the Abrikosov vortex lattice
requires lengthy and tedious calculations.
We will consider first the collision of an impurity
with a single vortex, and the transitions that ensue.
In order to simplify the analysis, we will study a two
dimensional version of the problem. The vortex defines the z axis.
For very anisotropic superconductors,
like many high-$T_c$ materials, a 2D picture is fully
appropiate. In general, core states give rise to bands
along the z direction. Transitions are characterized
by the wavevectors, $k_z, {k'}_z \sim k_F$, of the states
involved. If the impurity potential is sufficiently short
range, we can ignore the dependence of the transition rates
on these momenta. If we neglect as well the dispersion of these
one dimensional subbands, and describe each of them by a
typical energy, $\epsilon_i$, the problem becomes again two
dimensional. The analysis to be discussed below applies,
with the only inclusion of a prefactor $k_F a_z$, where
$a_z$ is the range of the impurity potential along
the z direction. In any case, the most dramatic deviations
from standard flux flow dissipation will be shown
to be originated by the existence of a sizeable
energy gap within the vortex core. This is a general feature
of superconductors with a short coherence length.

A  simplified sketch of the impurity vortex collision is
shown in the figure.
Within second order perturbation theory, the matrix
element of the impurity potential and the states
involved in the transition are required. The  quasiparticle
states induced at the vortex core have a smooth
envelope, with a characteristic localization length
of the order of the coherence length of
the superconductor, $\xi$, and rapidly varying
oscillations, characterized by $k_F^{-1}$, where $k_F$ is the Fermi
wavevector
\cite{deGennes2,Schluter,Yury}.
Thus, the spatial dependence of these wavefunctions, within a plane normal to
the vortex, is
like $\Psi_i \sim \xi^{-1} e^{- |\vec{r}| / \xi} e^{i k_F r}$.
In addition, $\Psi$ shows an angular dependence. Its influence in dissipative
processes, however, is much weaker than the modulations discussed above.

For convenience, we will assume that the range of the
(2D) impurity potential is shorter than the length scales
mentioned above. Then, we can neglect its $q$
dependence, and write it as $\sum_q V a^2 e^{i \vec{q} \vec{r}}$,
where $V$ is the strength of the potential, and $a$ is a length of the order
of the range of the impurity potential.
 We focus on a single impurity,
which follows a trajectory $\vec{r} = \vec{r_0} + \vec{v} t$.
The vortex is located at the origin of coordinates.
The transition amplitude between two vortex states $i,j$, with
an energy splitting $\epsilon_{i,j}$ is given,
within second order perturbation theory, by:

\begin{equation}
A \sim {{V a^2} \over{\hbar \xi^2}} \int_{-\infty}^{\infty}
e^{- | \vec{r} (t) | / \xi} e^{i k_F r ( t )}
e^{ {i \epsilon_{i,j} t}\over{ \hbar}} dt
\end{equation}

In terms of the transition amplitude $A$, the energy dissipation per unit time
and per unit vortex length can be written as:

\begin{equation}
{{\partial E}\over{\partial t}} \sim
\sum_{\epsilon_{i,j}} \epsilon_{i,j} n v \int | A ( r_0 ) |^2 d r_0
\end{equation}

where the summation is over all intravortex transitions,
and an integral over all impact parameters, $r_0$, is required.
This integral is cut off by the smallest separation, $r_{min}$,
possible between the impurity and the vortex center,
This distance is of the order of the lattice spacing,
which we can take similar to $k_F^{-1}$.
$n$ is the (3D) impurity concentration.

The integral over time in eq. (1)
is determined by times such that $t < t_{min} \hbar / \epsilon$,
where we are assuming that $t_{min}$ is the shortest time
in the integrand.
Its behavior, as function of $r_0$ shows two different regimes,
depending on the value of $\hbar v / ( \epsilon r_0 )$. If
$\hbar v / \epsilon \ll r_0$, the integral has an exponential dependence
on $\epsilon / ( \hbar v  \hbox{min} ( k_F^{-1} , \xi , r_0 )$.
Due to the existence of a short distance cutoff in the integral
over $r_0$, this can be the relevant regime for superconductors
with shoet coherence length, as discussed below.
In that case, the energy dissipation
can be written as:

\begin{equation}
{{\partial E}\over{\partial t}} \propto \sum_{\epsilon_{i,j}}
{{n \hbar V^2 a^4 r_{min}^2 v^2}\over{\epsilon_{i,j}^2 \xi^4
\hbox{min} ( k_F^{-1} , \xi )^2}}\exp \left[ - {{\epsilon_{i,j} \hbox{min}
( k_F^{-1} , \xi , r_{min} )}\over{\hbar v}} \right]
\end{equation}

It is worth noting that, on the other hand, equation (2) gives rise to the
standard
ohmic dissipation for a metallic density of excitations.
For an extended, normal vortex core, we can replace the
sum over possible transitions by an integral. The density of
low energy excitations in a normal metal goes as
$ N ( \epsilon ) \sim \epsilon / E_F^2 $. Inserting this expression
in equation (2), we find that $ \partial E / \partial t \sim v^2 $,
which leads to ohmic behavior.

Equations (2) and (3) give the energy dissipation due to a single vortex.
The dissipation per unit volume can be obtained by multiplying by the
(two dimensional) density of vortices, $\sim H / ( H_{c_2} \xi^2 )$,
where $H$ is the applied field.

In order to relate the dissipation to the I-V characteristics of
the material, we will assume that the voltage drop is due to the
vortex motion\cite{Parks}. Then, the electric field within the sample is
$| \vec{E} | = v | \vec{H} | / c$.  The dissipation per unit volume
is given by $\sigma_{ff} | \vec{E} |^2$, where $\sigma_{ff}$ is the flux flow
conductivity.

To our knowledge, there are no direct measurements of the vortex velocity
in a type-II superconductor.
For magnetic and electric fields typical for high-$T_c$ materials,
$H \sim 1$T, $E \sim 1$V/cm, we find that $v/c \sim 10^{-6} - 10^{-5}$.
This result implies that, when the lowest energy splitting within the vortex,
$\epsilon \sim \Delta / ( k_F \xi )$,
is $\epsilon \sim \hbar v / r_{min} \sim 2$meV, the effects
considered here become relevant (we are taking $r_{min} \sim 1$\AA).
This value is below the estimated gap of high-$T_c$ superconductors,
$\Delta \sim 3-5 k_B T_c$ ( assuming conventional s-wave pairing).
This estimate can be reduced for larger magnetic fields or smaller
voltages.

The dependence of the transition rates on $v$, for sufficiently
low values of $v$, is determined by the exponential in eq. (3).
Expressing the dissipation as $\vec{j} \vec{E}$, and using the
fact that $\vec{E} \propto \vec{v}$, we conclude that
$I \propto \exp - ( V_0 / V ) $,  where $V_0$ is a threshold
voltage. Its value is such that the local electric field is
given by
$ E_0 \sim \epsilon r_{min}  B / ( \hbar c )$.
For the parameters mentioned above, this field
is 1 V/cm.
For fields lower than
this value, dissipative processes are strongly suppressed.
Note that this effect should be more easily observable at
high magnetic fields, as the velocities of the vortices are
smaller, and the motion is less effective in inducing
transitions. This decrease in the vortex velocities
more than compensates the increase in the number of vortices induced
by the magnetic field. Thus, flux flow dissipation, for a fixed
voltage, should decrease with increasing magnetic field, in sharp contrast
with the Bardeen-Stephen model\cite{BS}.

The anomalous dissipation in superconductors with short
coherence length can be masked by pinning effects.
Pinning energies also show unconventional dependence
on applied currents, magnetic fields and temperature in high-$T_c$ materials.
The effects reported here influence the flux dynamics mostly
at low temperatures (short coherence lengths), which also enhance the pinning
of vortices.
High magnetic fields, on the other hand, reduce pinning, and require
lower vortex velocities to generate a given voltage drop. Thus,
we expect that the suppression of vortex viscosity
will be most effective in modifying the properties
of high-T$_c$ superconductors at low temperatures and high
magnetic fields.

In many situations, flux creep determines the magnetic properties.
The dynamics of vortices are described in terms of an
exponential dependence on an activation energy, and a characteristic
frequency\cite{Palstra},
which describes the motion of the vortex within each potential well.
This frequency, when the motion of the vortices is overdamped,
depends on the viscosity. So far, its dependence on magnetic field
and temperature has been neglected. The analysis reported here suggests
that this prefactor can also have a strong dependence on fields
and temperatures. As function of temperature, we expect a crossover
from a high temperature regime. where the motion is overdamped,
and this attempt frequency is determined by the viscosity alone,
to a low temperature regime, where the viscosity is negligible,
and the attempt frequency depends on the mass of the vortex.

Our results may help to explain anomalous effects in the
ac response of high-T$_c$ superconductors in megnetic fields\cite{Kes}.
The strong dependence of $I$ on $V$, for low voltages, described
above may help to explain a variety of experiments, which have
been fitted to $I \sim V^{\alpha}$ laws, with $\alpha \gg 1$\cite{Font}.
The absence of dissipative mechanisms, at low temperatures and
high magnetic fields may also suppress the magnetic
relaxation rate\cite{Crabtree}. Magnetic relaxation  involves
the exchange of energy from the magnetic field into
thermal energy.
Our estimates for the relevant vortex velocities
are also very close to those of interest in the vortex tunneling regime
\cite{Simanek}. A change in the dissipation mechanisms can induce
drastic modifications in the tunneling rates, as they depend
exponentially on the viscosity.

It is interesting to speculate with the behavior of a type II
superconductor in the limit of very large energy splittings
within the vortex cores. Then, conventional dissipative mechanisms
are frozen out, and, in the absence of pinning centers, vortices
will move without friction. The motion of vortices gives rise
to time dependent magnetic fluxes and induced electric fields,
as discussed above. Thus, a voltage drop accross the superconductor
will be generated. On the other hand, the absence of dissipation
impedes the existence of currents. Unlike a superconductor
withouth an applied magnetic field, such a system should
be able to sustain a voltage drop, without induced currents.
This unusual behavior, which
closely resembles that of an insulator, will be masked by the
existence of pinning, which can localize the vortices.

The analysis presented so far relies on the applicability
of the BCS theory to superconductors where $\xi \sim k_F^{-1}$.
The limit of a strongly coupled superconductor, however,
loosely resembles an array of weakly coupled superconductiong
islands. In these systems, local superconductivity can appear
at temperatures well above those  where global coherence sets in.
This is similar to the expected behavior of a strongly coupled
superconductor, where preexisting pairs are supposed to
exist above $T_c$. Dissipation at low temperatures in
arrays of Josephson junctions is being studied extensively,
both experimentally\cite{Mooij} and theoretically\cite{Josephson}.
It seems confirmed that, in the absence of nonsuperconducting,
resistive channels, dissipation is strongly reduced at low
temperatures. This result can be understood on general grounds,
since the superconducting system has no low energy excitations
below the plasma frequency.

In conclusion, we have extended the standard theory of flux flow
dissipation to superconductors where the level splittings, $\epsilon$, between
the states within the vortex core cannot be neglected ($\epsilon \sim \hbar v /
\xi$,
where $v$ is the velocity of the vortices). In this regime,
dissipation is strongly suppressed. The response of the superconductor
is non-ohmic, and $I$ shows a marked dependence on $V$. The
resistence is anomalously large. At a given voltage, the system undergoes
a crossover into a more conventional flux flow regime.

Simple estimates show that this analysis may be relevant to high-$T_c$
superconductors, at sufficiently low temperatures, so that $k_F \xi ( T ) \sim
1$.
The currents should be large enough, in order to avoid pinning
effects.
It would be desirable to explore this regime of temperatures, magnetic
fields and currents\cite{fluxflow}, in order to check the results outlined
here.

This work has been partially supported by CICyT (grant MAT91-0905).
We acknowledge illuminating discussions
on collision induced transitions and dissipation with
Prof. P. M. Echenique.

\figure{
Schematic picture of a vortex-impurity scattering process.
See text for details.}


\begin{references}

\bibitem{BS}
J. Bardeen and M. J. Stephen, Phys. Rev. B {\bf 140}, 1197 (1965).

\bibitem{NV}
P. Nozi\`eres and W. F. Vinen, Philos. Mag. {\bf 14}, 667 (1966).

\bibitem{deGennes}
P. G. de Gennes, {\it Superconductivity of metals and alloys},
Addison Wesley, Reading, MA (1989).

\bibitem{Schmid}
A. Schmid, Physik Kondens. Mat. {\bf 5}, 302 (1966).

\bibitem{Maki}
C. Caroli and K. Maki, Phys. Rev. {\bf 164}, 591 (1967).

\bibitem{deGennes2}
C. Caroli, G. de Gennes and J. Matricon, Phys. Lett. {\bf 9},
307 (1964); C. Caroli and J. Matricon,
Phys. Kondens. Mater. {\bf 3}, 380 (1965).

\bibitem{Schluter}
F. Gygi and M. Schl\"uter, Phys. Rev. B {\bf 43},
7609 (1991).

\bibitem{states}
K. Karrai, E. J. Choi, F. Dunmore, S. Liu, H. D. Drew, Qi Li, D. B. Fenner, Y.
D. Zhu and
F. C. Zhang, Phys. Rev.Lett. {\bf 69}, 152 (1992).

\bibitem{Ashcroft}
N. W. Ashcroft and N. D. Mermin, {\it Solid State Physics}, Holt-Saunders, New
York (1976).

\bibitem{Guinea}
F. Guinea, Phys. Rev. Lett. {\bf 53}, 1269 (1984).

\bibitem{Echenique}
A. Arnau, P. M. Echenique and R. H. Ritchie, Nucl. Inst. and Methods, {\bf
B33}, 138 (1988).

\bibitem{Yury}
Yu. Pogorelov, preprint.

\bibitem{Parks}
See, for instance, Y. B. Kim and M. J. Stephen in {\it Superconductivity}, R.
D.
Parks ed., M. Dekker, New York (1969).

\bibitem{Palstra}
T. T. M. Palstra, B. Battlogg, R. B. van Dover, L. F. Schneemeyer
and I. V. Waszczak, Phys. Rev. B {\bf 41}, 6621 (1990).

\bibitem{Kes}
F. Zuo, M. B. Salamon, E. D. Bukowski, J. P. Rice and D. M. Ginsberg,
Phys. Rev. B {\bf 41}, 6600 (1990); L. F\'abregas, J. Fontcuberta,S. Pi\~nol,
C. J. van
der Beek and P. H. Kes, Phys. Rev. B {\bf 47}, 15250 (1993).

\bibitem{Font}
E. Zeldov, N. M. Amer, G. Koren, A. Gupta, T. R. J. Gambino and M. W. Mc
Elfresh,
Phys. Rev. Lett. {\bf 62}, 3093 (1989).
M. A. Crusellas, J. Fontcuberta and S. Pi\~nol, Phys. Rev. B, in press.

\bibitem{Crabtree}
G. T. Seidler, C. S. Carrillo, T. F. Rosenbaum, U. Welp, G. W. Crabtree
and V. M. Vinokur, Phys. Rev. Lett. {\bf 70}, 2814 (1993).

\bibitem{Mooij}
H. S. J. van der Zant, F. C. Fritschy, T. P. Orlando and J. E. Mooij,
Phys. Rev. B {\bf 47}, 295 (1993).

\bibitem{Josephson}
U. Geigenm\"uller, C. J. Lobb and C. B. Whan, Phys. Rev. B {\bf 47}, 348
(1993);
U. Eckern and E. B. Sonin, {\it ibid} {\bf 47}, 505 (1993).

\bibitem{Simanek}
Theoretical estimates are given by E. Simanek, Phys. Rev. B {\bf 46},
14054 (1992). For experimental observations, see
A. C. Mota, A. Pollini, P. Visani, K. A. M\"uller
and J. G. Bednorz, Phys. Rev. B {\bf 36}, 4011 (1987),
A. Garc{\'\i}a, X. X. Zhang, A. M. Testa, D. Fiorani and J. Tejada,
J. Phys.: Condens. Matter. {\bf 4}, 10341 (1992).

\bibitem{fluxflow}
M. A. Kunchur, D. K. Kristen and J. M. Phillips, Phys. Rev. Lett. {\bf 70}, 998
(1993).

\end{references}
\end{document}